\documentclass[12pt,a4paper]{article}
\usepackage[left=2cm, right=2cm, top=3cm, bottom=3cm]{geometry}
\usepackage[latin1]{inputenc}
\usepackage{graphicx}
\usepackage{setspace}
\usepackage{amsfonts}
\usepackage{amssymb}
\usepackage{amsmath}

\title{ Harvesting graphics power for MD simulations}
\author{J.A. van Meel~\footnote{FOM Institute for Atomic and Molecular Physics,
Kruislaan 407, 1098 SJ Amsterdam, The Netherlands} ,
A. Arnold~\footnotemark[\value{footnote}] ,
D. Frenkel~\footnotemark[\value{footnote}] ,\\
S.F. Portegies Zwart \footnote{Astronomical Institute ''Anton Pannekoek'',
University of Amsterdam, Amsterdam, The Netherlands }
\footnote{Section Computational Science, University of Amsterdam, Amsterdam, The Netherlands} ,
R.G. Belleman\footnotemark[\value{footnote}] }

\begin{document}
\maketitle
\abstract{We discuss an implementation of molecular dynamics (MD) simulations on a
  graphic processing unit (GPU) in the NVIDIA CUDA language. We tested our code on a modern GPU, the
  NVIDIA GeForce 8800 GTX.  Results for two MD algorithms suitable for short-ranged and long-ranged
  interactions, and a congruential shift random number generator are presented. The performance of
  the GPU's is compared to their main processor counterpart. We achieve speedups of up to 80, 40
  and 150 fold, respectively.  With newest generation of GPU's one can run standard MD
  simulations at $10^7$ flops/\$.}

\section{Introduction}


Over the last 30 years computer simulations have become an important tool in materials science,
often bridging the gap between theory and experiment. Simulations can be used both to predict
the outcome of experiments and to test the assumptions of theories. The basic idea of most
classical simulations is to calculate the forces acting on all particles, and then integrate Newton's
equations of motion using these forces. This approach is not limited to single atoms or molecules,
the same approach can be used to model the motion of stars within galaxies.


With the rapid increase of available computational power, more systems become tractable for
simulations. Nowadays it is possible to simulate the time evolution of simple molecules over
microseconds with atomistic detail on a conventional personal computer.  However, for many
systems, the computational power of a single processor (CPU) is not sufficient. In this
case, simulations are run in parallel on many processors, which allows us to simulate hundreds
of thousands of molecules over time-spans of milliseconds. The increase of computational power
comes at a price: the different processors have to exchange information on the simulated system
continuously.  This communication costs time, reducing the effective
performance of a parallel system to typically less than 80\%
\cite{limbach:2006,lammps:2005} of the total
performance of all its individual CPUs. And although there are standardised software tools for the
implementation of this communication, such as PVM~\cite{sunderam:1990}, MPI~\cite{gabriel:2004}
 or OpenMP~\cite{dagum:1998}, writing a code for
parallel execution is not trivial. Moreover, the necessary very low latency, high
throughput communication hardware often costs as much as the processing units themselves.

An alternative approach speeds up the simulations by using special purpose hardware.
For example, in simulations of stars or charged molecules, more than 90\% of the computation time
is typically spent on the calculation of the gravitational or electrostatic interaction. Most
prominently, the GRAPE board~\cite{makino:1998} is a special purpose hardware
designed to calculate such
interactions; recently, a variant called MDGRAPE~\cite{susukita:2003} has been
put forward to calculate the interactions of more general pair potentials.
Due to their specificity, these boards can
achieve several orders of magnitude higher throughput compared to conventional
CPUs, but are only of interest for a limited community of
researchers. This makes these boards relatively expensive and their development cycle long.

Since 2003, a new route to gain additional computational power has
opened: the graphics processors (GPUs) of recent PC hardware have become general
purpose processors, which can be programmed using C--like programming
environments such as the GL shader language (GLSL)~\cite{rost:2004}, C for graphics
(Cg)~\cite{fernando:2003} or the NVIDIA compute unified device architecture
(CUDA)~\cite{nvidia:2006}. Their computational power exceeds that of the CPU by orders of magnitude: while
a conventional CPU has a peak performance of around 20 Gigaflops, a NVIDIA GeForce 8800 Ultra reaches
theoretically 500 Gigaflops. This means, that 4 graphics cards can replace a complete 64
processor PC cluster, saving space and reducing the necessary power supply from 15kW to around 2kW.
Moreover, graphics processors follow a Moore--law with a computational power doubling
every 9 months, in contrast to 18 months for conventional CPUs. For
the end of 2007, the first Teraflops-cards are expected.

There have been early attempts to harvest this computational power for
various applications, including fast Fourier transforms~\cite{larsen:2001}, matrix
operations~\cite{moreland:2003}, lattice Boltzmann simulations~\cite{li:2003} or
Monte Carlo (MC) simulations of the 2D Ising model~\cite{tomov:2005}. Recently,
Portegies Zwart et al.~\cite{portegies:2007, belleman:2007} presented a N-body simulation with
gravitational interactions, where the force calculation was performed on the
GPU.  For the latter application, the graphics cards are in direct competition
with the GRAPE boards, and achieve similar performances at much lower costs and
higher reliability. Yang et al. already presented a
proof--of--concept molecular dynamics simulation for the thermal conductivity of
solid argon~\cite{yang:2007}; their implementation is however limited to the
simulation of defect--free solids.

In this article, we aim to assess the portability of classical molecular simulation systems onto
GPUs using NVIDIA's CUDA~\cite{nvidia:2006}. Unlike the previous attempts of putting only the
computationally most expensive parts of the
simulation onto the graphics cards, we demonstrate that in fact the entire simulation can be
ported to the graphics cards. The resulting program reproduces all data obtained from a
standard single--processor simulation.
We report benchmarks of three codes: two simulating the classical
``work--horse'' of coarse--grained molecular
simulation, the Lennard-Jones system, and a classical rand48 random number
generator~\cite{unix:2002}. We tested these codes on a system consisting of an
Intel Xeon CPU running at 3.2 GHz and a NVIDIA GeForce 8800 GTX
(16 multiprocessors, running at 675 MHz each). For both the simulation and the
calculation of random numbers, we achieve an about 25-- to 150--times speedup
using the GPU compared to the CPU.

\subsection{GPU architecture}

\begin{figure}[htb]
  \centering
  \includegraphics[width=.6\textwidth]{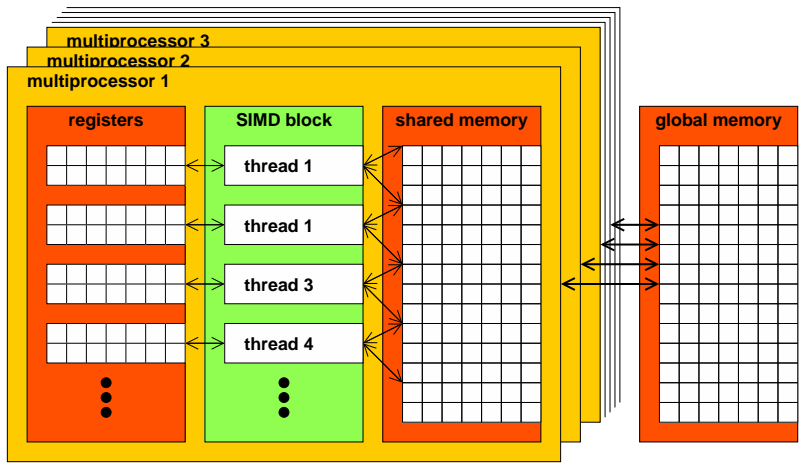}
  \caption{Schema of the multiprocessor and memory organisation of current NVIDIA
     GPUs. While registers are bound to a thread, all threads of the same SIMD block
    have access to a common shared memory, and all threads on all
    MPs have access to the global memory.}
  \label{fig:gpuarch}
\end{figure}

To facilitate the discussion on the technical implementations, it is necessary
to briefly summarise the key aspects of the GPUs hardware architecture and its
nomenclature (see also Figure~\ref{fig:gpuarch}). We use the NVIDIA CUDA system
for programming the GPU, which allows to write functions for the GPU, so--called
\emph{kernels}, in a C--like language. For detailed information we refer to
NVIDIA CUDA programming guide~\cite{nvidia:2006}.

The NVIDIA GeForce 8800 GTX consists of 16 multiprocessors (MPs). Each MP has a
single--instruction--multiple--data (SIMD) architecture and is capable of
performing 32 times the same operation on different data per two clock cycles.
Many copies of a kernel, so-called \emph{threads}, are executed in parallel on
all available MPs on the GPU. To fit the SIMD architecture, groups of 32 threads
form a \emph{warp} which is executed on the same MP. If a kernel contains a
branch and threads of the same warp take different routes, then both routes are
executed sequentially and the total run time is the sum of both branches. This
\emph{warp divergence} can have a serious impact on performance.

Threads can store data in 8192 32--bit registers per MP, and a high-speed
\emph{shared memory} of 16 KB per MP is available to share data among threads
running on the same MP. For this, threads are grouped into \emph{blocks} of up
to 512 threads which are forced to run on the same MP. A slower \emph{global
  memory} of 768 MB is also available that is shared among all MPs. To hide
register read-write latencies of one to two clock cycles, it is recommended to use
block sizes of 192 or more threads, and more than one block per MP should be
scheduled in order to hide the much larger global memory read latencies of 200
to 400 clock cycles. Note however, that GPU global memory is still ten times
faster than the main memory of recent PCs.

As a final remark we point out that nowadays graphics hardware only
supports single precision floating point arithmetic. This might not
suffice for systems where energy conservation is crucial. But for
systems in thermal equilibrium, i.e. with a stochastic thermostat,
this forms no limitation.

\section{N-squared MD}\label{sec:n2}
We start with the most simple molecular dynamics algorithm
in which each particle interacts with all other particles.
Therefore, the total force calculation scales quadratic with
the particle number $N$. The force $\vec{f}_i$ on a particle
$i$ is given by
\begin{equation}
\vec{f}_i = - \sum_{\substack{j = 1\\j \ne i}}^{N} f(|\vec{r}_j-\vec{r}_i|)\frac{\vec{r}_j-\vec{r}_i}{|\vec{r}_j-\vec{r}_i|}\,,
\label{eqn:n2_force}
\end{equation}
where $f(r)$ is the well-known Lennard-Jones pair force,
truncated at a distance $r_c = 2.5\sigma$ and shifted such that
the force at the cutoff distance was zero. The full Lennard-Jones pair
force is given by
\begin{equation}
  f_\text{LJ}(r) = 24\epsilon \left[ 2\left(\frac{\sigma}{r}\right)^{13}
    - \left(\frac{\sigma}{r}\right)^{7}\right],
\end{equation}
our truncated and shifted force by
\begin{equation}
  f(r) = \begin{cases}
    f_\text{LJ}(r) - f_\text{LJ}(r_c) & r<r_c \\
    0 & r\ge r_c\,.
    \end{cases}
\end{equation}
The Velocity Verlet algorithm was applied to integrate Newton's equations of motion
\cite{frenkel:2002}.

\subsection{Implementation details}
\begin{figure}[htb]
\center
\includegraphics[width=0.3\textwidth]{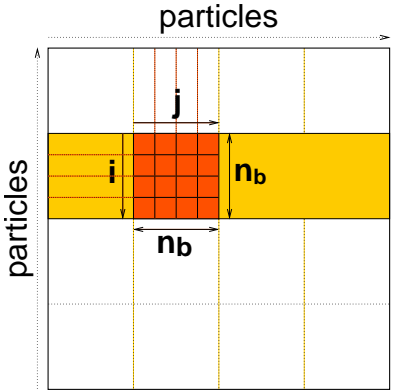}
\caption{Schema of the splitting of the force calculation into blocks. Each
  of the $n_b$ threads is dedicated to calculating all interactions of particle
  $i$. All particles $j$ are loaded in blocks of again $n_b$ particles, which
  are then shared among all threads.}
\label{fig:n2_scheme}
\end{figure}
A Molecular Dynamics simulation is naturally suited for a SIMD architecture,
because it performs the same set of operations on each particle.
The most simple way to parallelise this algorithm is to have one
independent thread per particle. However, naively implementing Equation
\ref{eqn:n2_force} turns out to be far from efficient. The reason for this is
that every thread loads all particle positions from global memory, which is not cached.
Each read access comes with some latency causing the processor to idle until
the data arrives. A huge improvement can be achieved by taking advantage
of the fact that all threads need the same data. By grouping threads into
blocks, data can be shared among them, effectively reducing memory bandwidth
and idle times.

Our implementation works as follows: each thread loads one different particle
from global memory and stores it into shared memory. Then all threads of a block
are synchronised to ensure loading has
finished. Now the data of all threads are accessible through high--speed
shared memory, and each thread can calculate the interactions of its dedicated
particle with all other particles in shared memory (see figure~\ref{fig:n2_scheme}).
For a block of $n_B$ threads, this reduces memory bandwidth by a factor $1 / n_B$. In
addition, each thread can now compute more interactions per memory read, allowing
the thread scheduler to more efficiently hide global memory latencies.
The optimal block size depends on the resources used by the kernel: number of
registers and shared memory size. A block size of $n_B = 64$ turned out to be
the optimal choice for our program. For details about the interplay
between register usage, shared memory usage, block size and number
of blocks per multiprocessor, we refer to NVIDIA's CUDA programming
guide \cite{nvidia:2006}.

\subsection{Results}
\begin{figure}[htb]
\center
\includegraphics[width=0.45\textwidth]{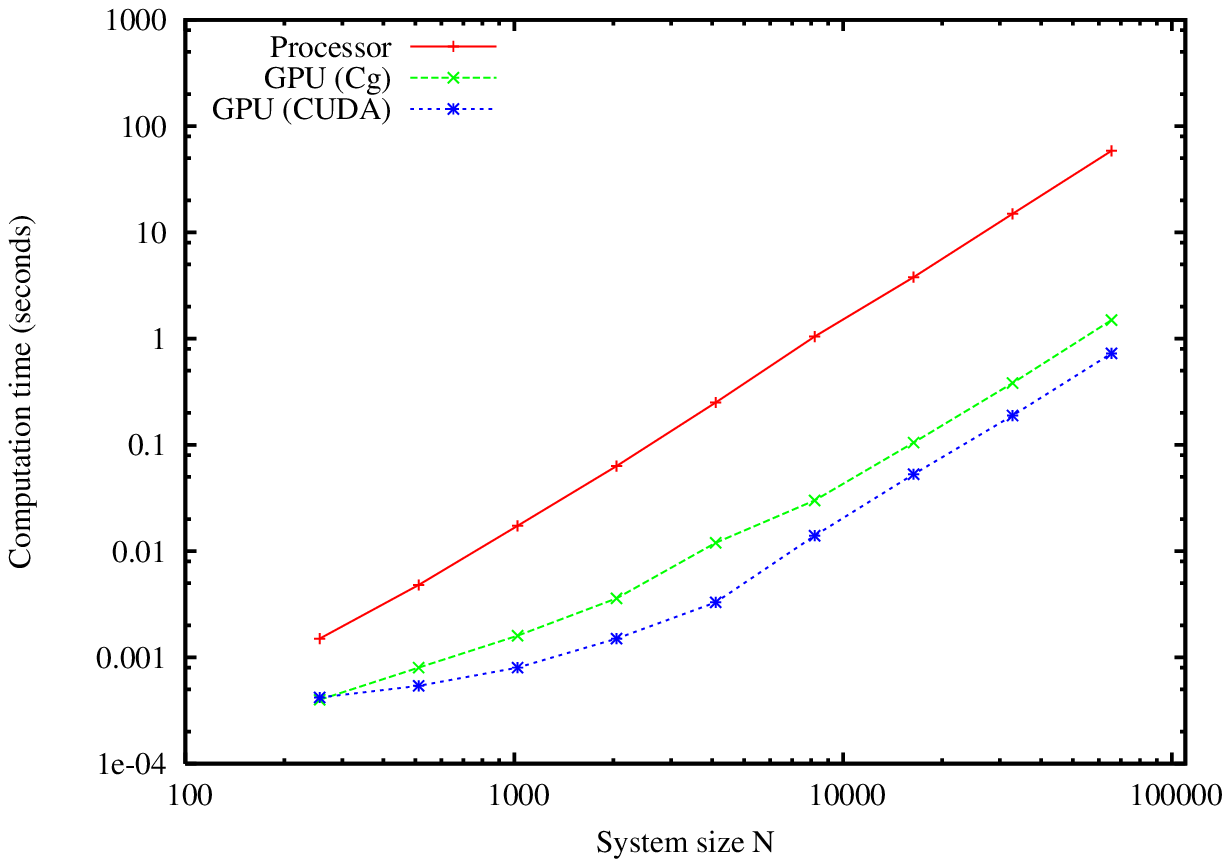}
\includegraphics[width=0.45\textwidth]{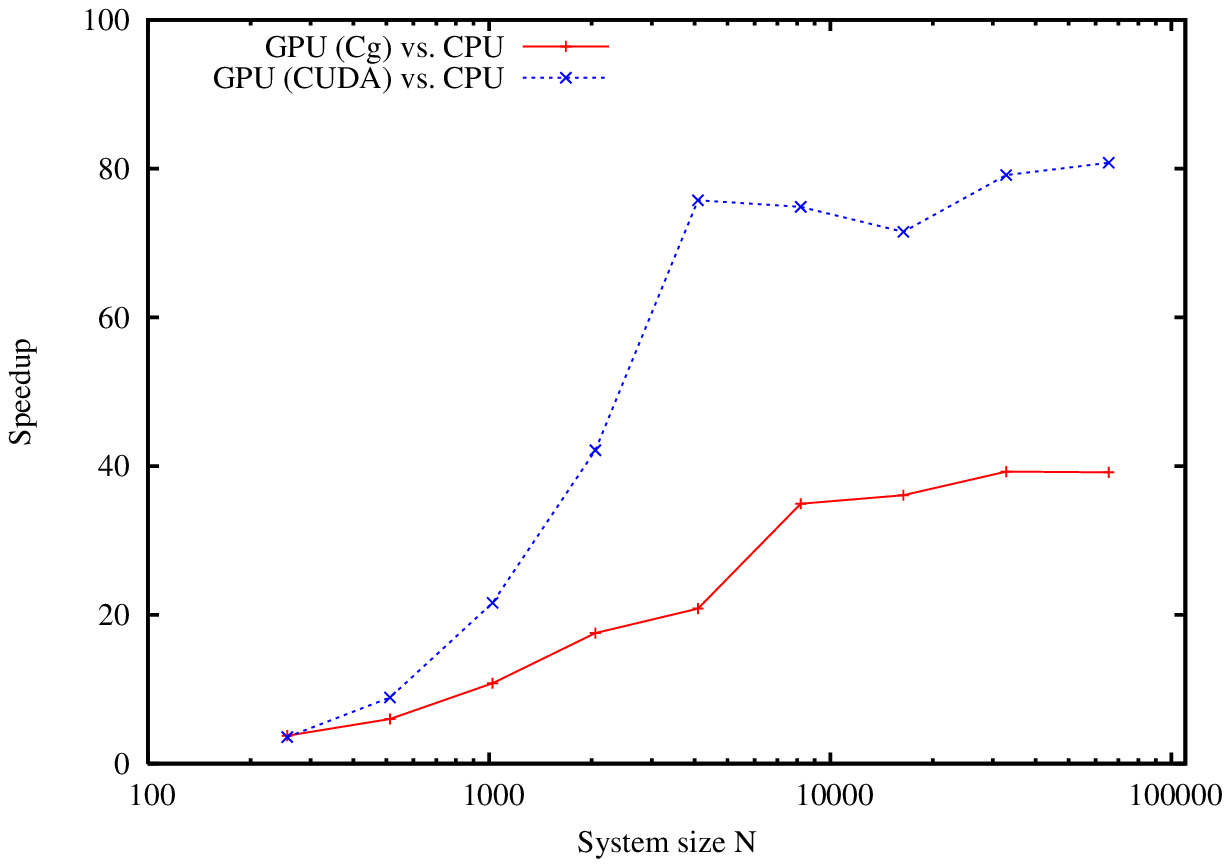}
\caption{Left: Time (in seconds) required to integrate a single MD time
step as a function of system size $N$. Right: Speedup factor for both GPU
implementations, Cg and CUDA. The speedup saturates once both the processor
and GPU versions have reached the quadratic scaling regime.}
\label{fig:n2}
\end{figure}
To compare the performance of the GPU and CPU implementations on our test system,
we measured the time required to integrate a single MD time step
as a function of system size $N$. The speedup factor $x$ is defined
as
\begin{equation}
x = \frac{ T_{CPU} } { T_{GPU} },
\end{equation}
where $T_{CPU}$ is the time used by the CPU implementation and
$T_{GPU}$ the time used by the GPU implementation. In addition
to the GPU implementation with CUDA, we also present results from
a GPU implementation in Cg \cite{fernando:2003}, a language designed for
graphic processing. It is supported by the majority of current
graphics hardware, but does not provide the flexibility required
for more complex MD algorithms.

The left graph in Figure \ref{fig:n2} shows the average time
used to integrate a single MD time step. The quadratic scaling
of run time with system size is clearly visible for the CPU
version. For the GPU code at small system sizes, the overhead of
invoking the graphic program is comparable to the actual
computation time. Therefore, the quadratic scaling regime is
reached when this overhead becomes negligible, which corresponds
to a system size of approximately 4000 particles.

The speedup factor for the GPU implementation is depicted in the
right graph in Figure \ref{fig:n2}. Although the GPU version is
faster for all our system sizes, it requires a system size larger
than 4000 particles to reach its full speedup of around 80.

\section{Cell-lists MD}\label{sec:cl}
If the pair interaction is short-ranged, the simulation
box is typically decomposed into smaller domains, so-called cells,
with a side length equal to or greater than the  maximum
interaction range. For a given particle, all interaction
partners are then located in the same and directly
neighbouring cells. Therefore, the algorithm scales linearly
with the number of particles, but suffers some penalty
due to the overhead associated with maintaining the cell structure.
For small systems, this might be disadvantageous compared
to the N-squared algorithm, but for large systems it generally
results in a huge performance gain. The system size at which
both algorithms perform equally well is called \emph{break--even
point}.

Another way of optimisation are the so-called Verlet
lists. For each particle, a list holds all neighbour particles
within a sphere of $r_{V} = r_c + \Delta r$, the Verlet radius.
The Verlet lists \emph{skin} with width $\Delta r$ prevents particles
to move into interaction range unnoticed and generally acts as an
invalidation criterion. Every time a particles list
is updated, the particles current position is stored as Verlet list
centre. If this particle moved further than $\Delta r / 2$ away from
its Verlet list centre, the list has expired and needs to be rebuild.
Obviously, the larger $\Delta r$, the less frequent the lists have to
be updated, but the more unnecessary interactions with $r > r_c$ have
to be computed. Updating Verlet lists is rather expensive and scales
like $O(N^2)$. Therefore, cell lists are often used to reduce its
costs to $O(N)$. Compared to cell lists, Verlet lists further reduce
the number of possible interaction partners and result in a theoretical
seven fold speedup.

Yang et al. \cite{yang:2007} used the Verlet lists approach to compute
the thermal conductivities in solid argon on a GPU. However, their Verlet
lists were computed only once (on the CPU) and never updated, which restricts
its use to defect--free solids. To fit the SIMD architecture, they added
virtual particles to obtain the same list size for all particles. Moreover,
to avoid inner-loop branching which deteriorates the performance, the
interaction cutoff distance
was set to the Verlet list radius $r_V$. In doing so, they removed the
essential skin from their Verlet lists allowing interactions due to
fluctuations to be ignored. This shows that Verlet lists are not
particularly suited for a SIMD architecture. The MD code of Ref.
\cite{yang:2007} is therefore useful as a proof of concept, but cannot be
used for production runs.

In our program we applied only cell lists. They seem more suitable for the
hardwares architecture and could be implemented to run entirely on the GPU.
Care was taken not to neglect any interactions and to include cell list updates.

\subsection{Implementation details}
There are plenty of schemes to implement the cell lists technique~\cite{frenkel:2002}.
One approach uses one linked list per cell to store the identities of the particles
located in it. The advantage is that this scheme works well for all
densities without parameter modifications, because there are no size limitations
on a linked list. The disadvantage is that memory access is random,
not sequential, and therefore a linked list cannot be loaded in parallel.

Another way is to assign a fixed sized array of placeholders (AOP)
to every cell and physically copy particles position into this array.
The advantage of this scheme is that interacting particles are
physically close together in memory allowing for fast parallel loading.
The disadvantage is that it generally requires more memory, because
each AOP has to provide space enough to store particles at the highest
possible density.

Our implementation uses the latter scheme. Per cell, one thread is
devoted to one placeholder. Empty places are filled with virtual particles.
Each thread $i$ of a cell $c_0$ loads the data of placeholder $i$ of cell $c_n$
from global memory and stores it into shared memory. It synchronises with
the other threads of the same cell $c_0$ to ensure loading has finished.
Now it computes the interactions of particle $i$ with all particles
in shared memory. Note that this is done for virtual particles, too.
These steps are performed for the centre cell, $c_n = c_0$, and all neighbour
cells, $n = 1 \dots 26$.

%
%

But the force computation is not the only task. As particles
move, the cell lists have to be updated. While this is straight
forward on a single CPU, the parallel version comes with some
difficulties. If a cell realises that one of its particles is about
to move to a neighbour cell, it cannot move the particle there without
the risk of memory--write conflicts and data inconsistency.

To safely update a cell in a parallel environment, we first remove
all particles from the list which left the cell. Then all particles
from neighbouring cells are checked to see if they moved in and need to
be added to this list. Double--buffering ensures that all old lists
stay intact until all cells have updated their list. Both for removing
the particles from a cell that have left the cell and adding the
particles that have moved in, we have to first test where a particle
belongs, and then update the corresponding particle list by either
deleting or adding particles.

Testing particles can be done in parallel. Each thread of a cell
computes the cell id for one particle and stores it in shared
memory. Now one thread sequentially loops over these particles
and adds those with a correct cell id to the list. To prevent
memory--write conflicts, this task has to be performed by a single
thread per cell, leaving all other threads idle.

Updating a cell list requires all particles from this cell and its
neighbour cells to be loaded from memory. In order not to load the
same data twice, we perform this task during force calculation,
not directly after the integration of positions. As a drawback the
cell lists are not precisely up--to--date, but one time-step behind.
In order not to neglect any interactions, the cells have to 
have a side length larger than the maximum interaction range plus
a so-called skin of thickness $\lambda$, where $\lambda$ is the
maximum particle displacement per time step. For MD simulations it
is common practise to use an even larger skin and therefore update
the cell lists only every couple of time steps.

\subsection{Results}
\begin{figure}[htb]
\center
\includegraphics[width=0.45\textwidth]{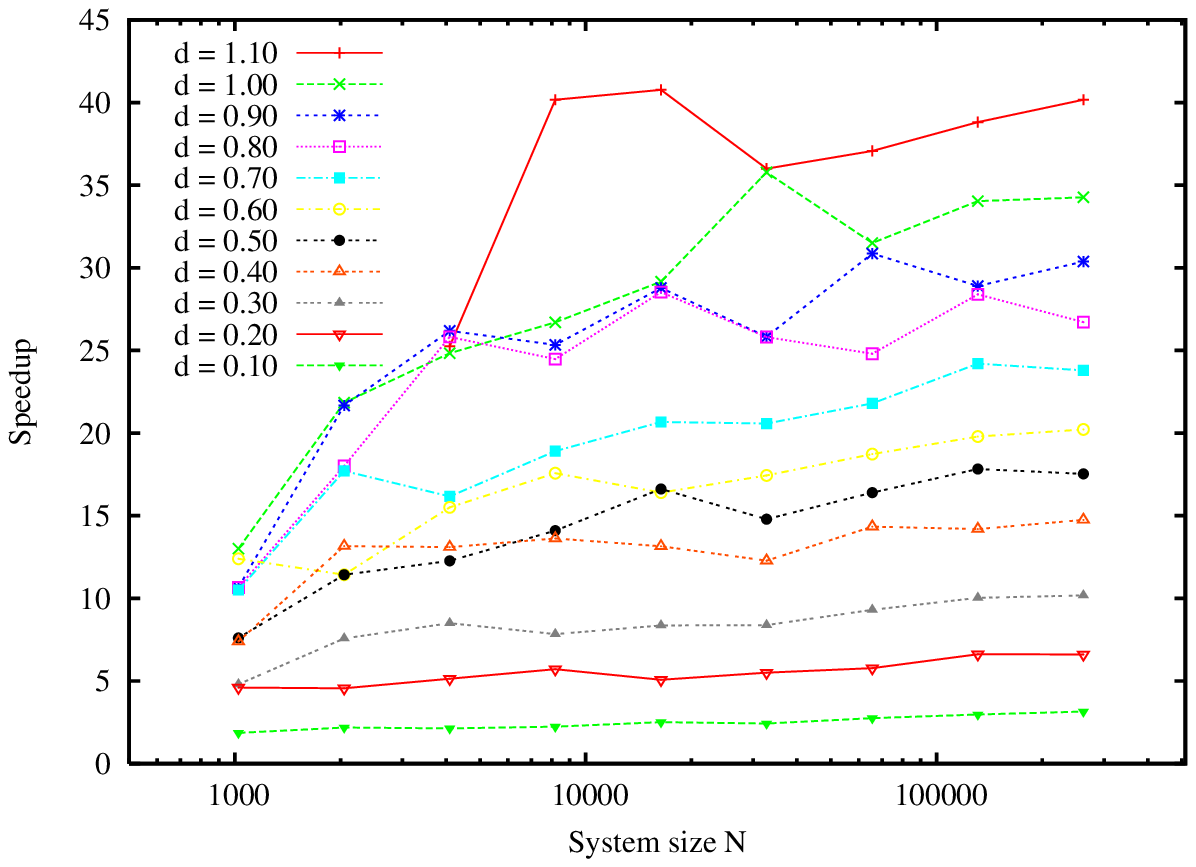}
\includegraphics[width=0.45\textwidth]{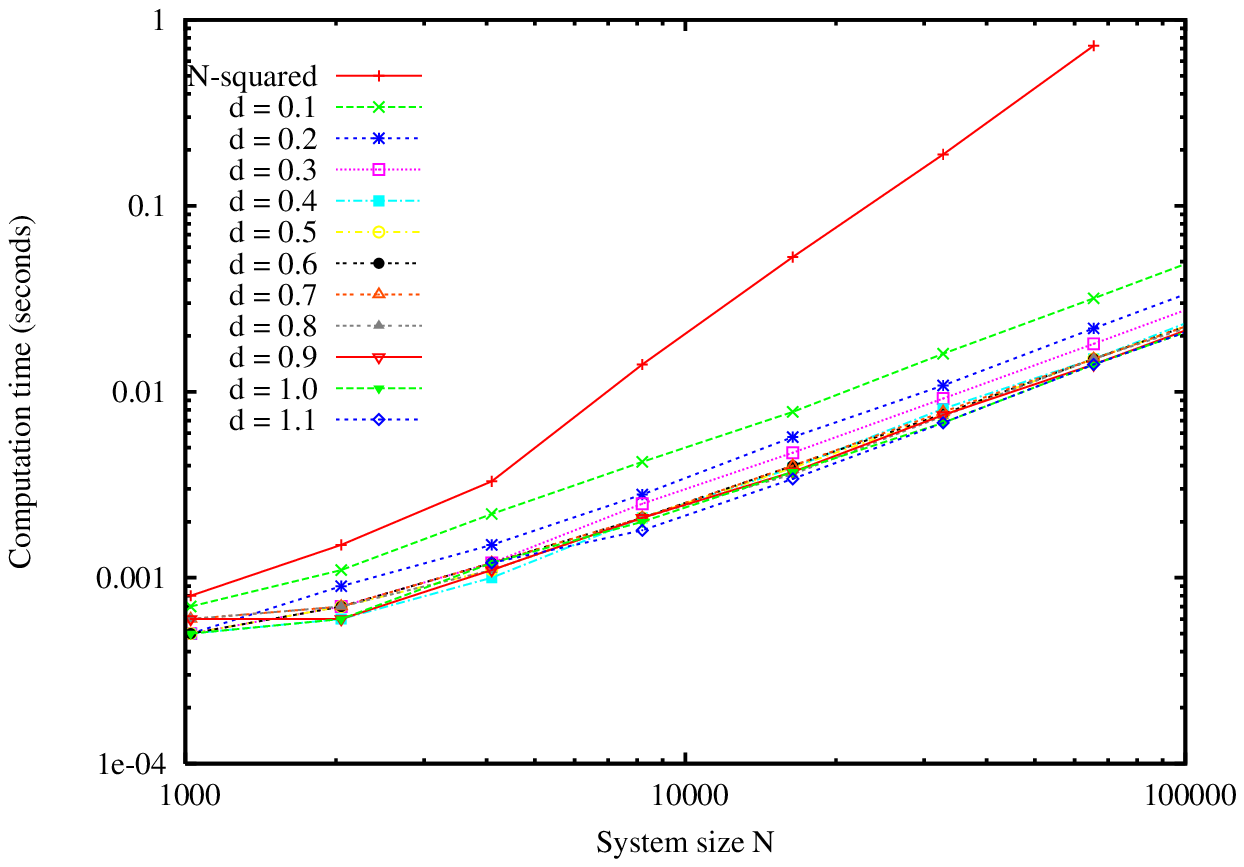}
\caption{Left: Speedup of the GPU version at various densities. The kinks
and bumps are reproducible. For details, see main text. Right: absolute times
(in seconds) for a single MD step. The density-independent N-squared
MD data is presented as a reference. The intersection points with this
reference data would indicate the break-even points for the cell-lists
algorithm. But for all system sizes shown here, the cell-lists MD is
faster than its N-squared counterpart.}
\label{fig:cl}
\end{figure}
As for the N-squared MD algorithm in Section \ref{sec:n2}, we compared the
GPU implementation with its CPU counterpart. Because the algorithmic
complexity exceeded the capabilities of Cg, only CUDA could be used.

The system size lower limit is given by the requirement to have at least $3$
cells per dimension at a density of $\rho = 1.0$. The upper limit for the
density was given by the array size associated with every cell, which was
$n = 32$ in the data presented here. For a minimum cell size of $r_c = 2.5 \sigma$
and a density of $\rho = 1.0$, on average $\rho \, r_c^3 \approx 16$ places per cell
are occupied. This implies
that most of the time half of the threads are calculating interactions of
virtual particles which do not contribute. This deteriorates at lower
densities.

The run time is density dependent: the more particles per cell, the
more interactions have to be computed, and the computation time rises.
This is true for the CPU version. However, our GPU version behaves
differently. In contrast to the CPU version, the GPU versions run time
is dominated by the total number of cells, not by the number of
interactions per cell. This is because interactions are always
calculated for all placeholders; at low densities, most of them are
however empty. At constant number of particles, the number of cells
decreases with the density, and therefore the run time decreases. This
effect saturates once all placeholders of a cell are used.

The left graph of Figure \ref{fig:cl} shows the speedup factor for our GPU
implementation. At the lowest density of $\rho = 0.1$, the GPU version is twice
as fast as the CPU version. At higher densities, the GPU outperforms the CPU
by up to a factor $40$. The errors for these speedup
factor are smaller than the symbol sizes and the kinks and bumps reproducible. They
relate to the cell size, which fluctuates in order to get an integer number of
cells per dimension. Assume a box length of $L_x = 11 \sigma$ and a minimum cell
size of $r_c = 2.5 \sigma$; then the number of cells for this dimension is
$n_x = \text{int}[ L_x / r_c ] = \text{int}[ 4.4 ] = 4$ and the actual cell size is
$r_c' = L_x / n_x = 2.75 \sigma$. This increase of $10$\% results into $33$\% 
more particles per cell, leading to $77$\% more interactions, decreasing the
CPU performance. But for the GPU version a few more threads compute real
particle interactions instead of virtual ones, resulting in no penalty.

The absolute computation times required per MD time step are depicted in
the right graph of Figure \ref{fig:cl}. For comparison, the (density
independent) N-squared MD data is shown as well. The cell-list data feature
a different slope than the N-squared data, indicating linear and quadratic
scaling, respectively. Intersection points with the N-squared curve would
indicate the break even points, where both algorithms perform equally well.
However, for all system sizes and densities depicted in Figure \ref{fig:cl},
the cell-lists version performed better than its N-squared counterpart.

\section{Random number generation}
For many applications in computer simulations, e.~g. Monte Carlo simulations
or Molecular Dynamics simulations with a stochastic thermostat, a large quantity
of (pseudo--)random numbers is required. Typically, simple linear congruential
generators such as the {\tt lrand48} are used~\cite{unix:2002}.
Given a number $x_n$, the following
number in its series is generated as follows:
\begin{equation}
  \label{eq:lrand48}
  x_{n+1} \equiv a x_n + c \mod 2^{48}
\end{equation}
where $a$ and $c$ are some integer constants. The pseudo--random number
$x_{n+1}$ is then converted to the pseudo--random number $Y_{n+1}$ of the
required data type. For $Y_{n+1}$, one uses the $d$ most--significant bits of
$x_{n+1}$, where $d$ is the bit size of the required data type (e.~g.  $d=31$
for nonnegative 32--bit integers).

\subsection{Implementation details}
To parallelise generation rule \ref{eq:lrand48}, we note that
\begin{equation}
  \label{eq:lrand48m}
  x_{n+m} \equiv A x_n + C \mod 2^{48}
\end{equation}
where
\begin{equation}
  \label{eq:AC}
  A\equiv a^m \mod 2^{48},\quad\text{and}\quad
  C\equiv \sum_{i=0}^m a^i c \mod 2^{48}.
\end{equation}

The random number generator is then implemented as follows. We choose a number
$S$ of random numbers to generate in parallel. To start the random number
generator, we choose a seed $x_0$, and generate the first $x_i$, $i=1\hdots S$
according to the serial rule \eqref{eq:lrand48}. The next set of $S$
pseudo--random numbers is then generated from this set according to rule
\eqref{eq:lrand48m}. Since the calculation of $x_{i+S}$ only requires knowledge
about $x_i$, all $x_{i+S}$, $i=1\hdots S$ can be calculated in parallel. If some
multiple $RS$ of $S$ random numbers is required, each set $x_i$, $x_{i+S}$,
$\hdots$, $x_{i+(R-1)S}$ can be calculated independently by an isolated
processor.

For the implementation on current GPUs, this is very convenient: Using $S$
independent threads, each thread first loads its current state $x_i$. Then, it
generates the following pseudo--random numbers $x_{i+S}$, calculates the output
value $Y_{i+S}$ and stores it. This step is repeated for all $x_{i+nS}$, until
in total $R$ random numbers have been generated and stored. Finally, the thread
saves the current state $x_{i+(R-1)S}$. For our test, we used $S=6144$
independent threads, grouped into 32 blocks of 192 threads each.

Note that all arithmetics is done modulo $2^{48}$. However, GPUs (as well as
standard CPUs) do not offer 48--bit data types. In principle, a 48--bit number
can be represented by three 16--bit numbers, but for performance reasons it is
better to represent the 48--bit number as one 64--bit integer or two 32--bit
integers. We have chosen to represent the $x_n$ by two 32--bit integers, which
contain the 24 most--significant and 24 lowest--significant bits.

\subsection{Results}

\begin{figure}[htb]
  \centering
  \includegraphics[width=.6\textwidth]{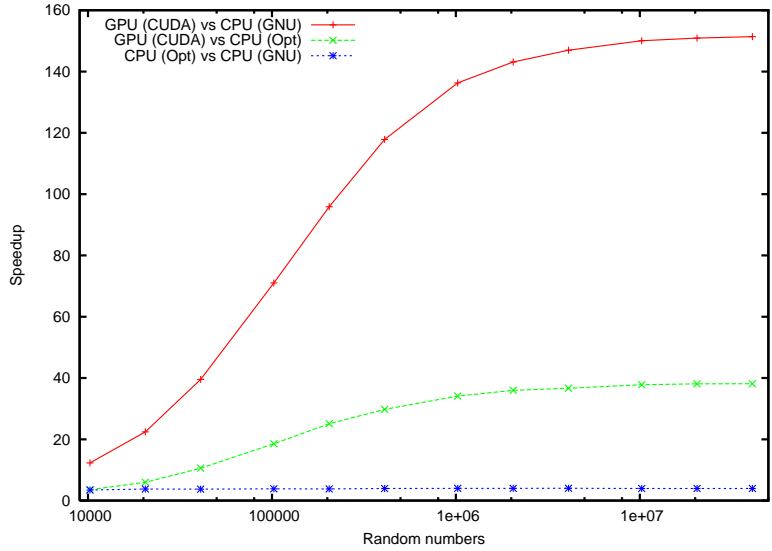}
  \caption{Speedup of a self-written CPU and a GPU version of the {\tt rand48}
    random number generator versus the standard {\tt glibc} implementation.}
  \label{fig:rng_ratio}
\end{figure}

We compare our implementation of the {\tt lrand48} random number generator on
the GPU both to the standard GNU--libc {\tt lrand48()} function as well as a
self--written CPU version using 64--bit arithmetics. For each
implementation, we measured the time necessary to generate the first $N$ random
numbers of the {\tt lrand48} series for $10,240\le N \le 40,960,000$. The resulting
speedup factors of our implementations relative to the GNU--libc implementations
are shown in Figure~\ref{fig:rng_ratio}.

The optimised CPU version is consistently faster than the system implementation by
a factor of almost four. This simply demonstrates the high 64--bit performance
of current PC processors. However, the GPU achieves a much higher
performance. For generating more than a million random numbers, the GPU is
faster than the standard--libc {\tt lrand48} by a factor of 150. Compared
to our optimised CPU--version, the speedup is still almost 40.

Although the speedup factor for this pure integer arithmetics problem is
therefore not as high as for typical floating--point problems, the GPU is still
competitive for the generation of random numbers. Moreover, our implementation
stores the output random numbers in the relatively slow main memory of the
graphics card. Depending on the problem at hand, it is however often possible to
generate random numbers on the fly, which will increase the speedup factor.

\section{Summary and outlook}


The computational power of recent graphics cards is fifty times as large as the
power of a conventional processor. It has been shown previously that this speed
can be harvested for many problems, e.~g. matrix multiplication or the
calculation of electrostatic interactions. In this article, we have demonstrated
that it is possible to run a conventional MD simulation entirely on a graphics
card. The simulations run 25-80 times faster than on a single conventional
processor, at comparable prices. This shows that it is also possible to harvest
this computational power for MD simulations.


Although our code features only the simple Lennard--Jones
potential, it is trivial to replace this potential by other pair-potentials,
including the coulombic interaction. By this, our code can in fact be used for
many systems of interest. Moreover, the GPUs are indeed general purpose
processors by now, and therefore it should be possible to implement many other
techniques equally efficiently, such as Ewald summation methods or SHAKE for
constrained dynamics. Currently the GPUs are limited to single precision
floating point operations. For long non--thermalized simulations, in which energy
conservation is crucial, this precision might not be sufficient. However, 
double--precision GPUs are expected for the end of this year.


While MD simulation techniques can be easily ported onto the GPU architecture,
this does not hold for the equally wide--spread family of Monte--Carlo methods.
Tomov et al.~\cite{tomov:2005} have implemented a MC scheme for the 2D Ising
model showing that lattice-based probabilistic simulations can be ported to
the GPUs SIMD architecture. However, off-lattice many particle MC simulations
are difficult to parallelise, both on conventional parallel architectures and
on SIMD hardware. Reasons for this are the random acceptance moves causing
unpredictable branching, and the permanent access to global information to
obey detailed balance.

The difference in computational power between conventional processors and GPUs
is expected to increase further. At the end of this year, NVIDIA GPUs are
expected to reach Teraflops performance on a single card, and will feature double
precision floating point operations, at a rate of 250 Gigaflops. Even with
current off--the--shelf PC mainboards it is possible to build systems
equipped with four graphics cards. A single PC can therefore obtain Teraflops
performance, and a small cluster of such PCs provides a computational power of
10 Teraflops for a price of less than \$100,000.

\section*{Acknowledgements}
This work is part of the research program of the Stichting voor Fundamenteel
Onderzoek der Materie (FOM), which is supported by the Nederlandse Organisatie
voor Wetenschappelijk Onderzoek (NWO). AA acknowledges support from the
Marie-Curie programme of the European Commission. SPZ and RB acknowledge
support by NWO (via grant \#635.000.303 and \#643.200.503) and the Netherlands
Advanced School for Astrophysics (NOVA).

\addcontentsline{toc}{section}{Bibliography}
\bibliographystyle{unsrt}   

\end{document}